\PassOptionsToPackage{unicode}{hyperref}
\PassOptionsToPackage{hyphens}{url}
\documentclass[
  conference]{IEEEtran}
\usepackage{xcolor}
\usepackage{amsmath,amssymb}
\usepackage{iftex}
\ifPDFTeX
  \usepackage[T1]{fontenc}
  \usepackage[utf8]{inputenc}
  \usepackage{textcomp} 
\else 
  \usepackage{unicode-math} 
  \defaultfontfeatures{Scale=MatchLowercase}
  \defaultfontfeatures[\rmfamily]{Ligatures=TeX,Scale=1}
\fi
\usepackage{lmodern}
\ifPDFTeX\else
\fi
\IfFileExists{upquote.sty}{\usepackage{upquote}}{}
\IfFileExists{microtype.sty}{
  \usepackage[]{microtype}
  \UseMicrotypeSet[protrusion]{basicmath} 
}{}
\makeatletter
\@ifundefined{KOMAClassName}{
  \IfFileExists{parskip.sty}{%
    \usepackage{parskip}
  }{
    \setlength{\parindent}{0pt}
    \setlength{\parskip}{6pt plus 2pt minus 1pt}}
}{
  \KOMAoptions{parskip=half}}
\makeatother
\usepackage{graphicx}
\makeatletter
\newsavebox\pandoc@box
\newcommand*\pandocbounded[1]{
  \sbox\pandoc@box{#1}%
  \Gscale@div\@tempa{\textheight}{\dimexpr\ht\pandoc@box+\dp\pandoc@box\relax}%
  \Gscale@div\@tempb{\linewidth}{\wd\pandoc@box}%
  \ifdim\@tempb\p@<\@tempa\p@\let\@tempa\@tempb\fi
  \ifdim\@tempa\p@<\p@\scalebox{\@tempa}{\usebox\pandoc@box}%
  \else\usebox{\pandoc@box}%
  \fi%
}
\def\fps@figure{htbp}
\makeatother
\usepackage{svg}
\NewDocumentCommand\citeproctext{}{}
\NewDocumentCommand\citeproc{mm}{%
  \begingroup\def\citeproctext{#2}\cite{#1}\endgroup}
\makeatletter
 \let\@cite@ofmt\@firstofone
 \def\@biblabel#1{}
 \def\@cite#1#2{{#1\if@tempswa , #2\fi}}
\makeatother
\newlength{\cslhangindent}
\newlength{\csllabelwidth}
\newenvironment{CSLReferences}[2] 
 {\begin{list}{}{%
  \setlength{\itemindent}{0pt}
  \setlength{\leftmargin}{0pt}
  \setlength{\parsep}{0pt}
  \ifodd #1
   \setlength{\leftmargin}{\cslhangindent}
   \setlength{\itemindent}{-1\cslhangindent}
  \fi
  \setlength{\itemsep}{#2\baselineskip}}}
 {\end{list}}
\usepackage{calc}

\providecommand{\tightlist}{%
  \setlength{\itemsep}{0pt}\setlength{\parskip}{0pt}}
\usepackage{booktabs}
\usepackage{array}
\usepackage{hyperref}
\usepackage{bookmark}
\IfFileExists{xurl.sty}{\usepackage{xurl}}{} 
\makeatletter
\@ifundefined{xmpquote}{}{}
\makeatother
\hypersetup{
  pdftitle={Structural Estimation of Marketing Mix Model Parameters from Geo-Experiments},
  pdfauthor={Niklas Heusch (niklas.heusch@zalando.de)},
  hidelinks,
  pdfcreator={LaTeX via pandoc}}

\title{Structural Estimation of Marketing Mix Model Parameters from
Geo-Experiments}
\author{Niklas Heusch (niklas.heusch@zalando.de)}
\date{Draft - September 2026}

\begin{document}
\maketitle
\begin{abstract}
Marketing Mix Models (MMMs) are widely used for marketing measurement
and budget allocation, but face fundamental identification challenges:
due to endogenous marketing spend decisions, MMM estimation on
observational time-series data cannot recover the true causal effects of
marketing. On the other hand, geo-experiments provide causal
identification through randomization, but it is not clear how to use
them efficiently to calibrate marketing mix models. We propose a novel
structural estimation approach that recovers the complete set of MMM
parameters---adstock decay (\(\alpha\)), saturation (\(\lambda\)), and
effectiveness (\(\beta\))---directly from geo-experimental time-series.
By differencing outcomes between treatment and control regions, our
method eliminates observed and unobserved confounding factors while
preserving the temporal variation that identifies each parameter. We
demonstrate on synthetic data that this approach recovers the true ROAS
and the response curve over the range of spending covered by the
experiments, together with credible estimates of the underlying
parameters. Our framework enables efficient pooling across multiple
experiments and provides a principled foundation for MMM calibration
that fully utilizes the information experiments contain.
\end{abstract}

\textbf{Companion materials:}
\href{https://github.com/niklas-heusch/mmm-materials}{Dataset} ·
\href{https://niklas-heusch.github.io/posts/mmm3-calibration/mmm3-calibration.html}{Illustrated
walkthrough}

\section{Introduction}\label{introduction}

Marketing Mix Modeling has become a standard approach for measuring
marketing effectiveness and guiding decisions on budget allocation. As
privacy regulations limit user-level tracking and third-party cookies
disappear, aggregate time-series approaches like MMM have gained renewed
importance (\citeproc{ref-jin2017bayesian}{Jin et al. 2017}). Several
widely-used software packages, including Google's Meridian, Meta's
Robyn, and pymc-marketing, implement MMM estimation, making these
methods accessible to practitioners across industries.

However, MMM faces a fundamental identification challenge: marketing
budgets are not randomly assigned. Companies increase advertising during
periods of expected high demand, algorithmic bidding systems chase
performance signals, and strategic decisions coordinate marketing with
promotional calendars. This endogeneity means that standard time-series
estimation cannot distinguish whether sales increased because of
marketing or whether marketing increased because sales were expected to
be high. In other words, MMMs struggle to estimate the effects of
marketing, as - in the presence of unobserved confounders - they cannot
tell causality apart from correlation.

Geo-experiments provide a solution through randomization. By randomly
assigning geographic regions to treatment and control conditions, these
experiments break the endogenous relationship between marketing and
demand, enabling causal identification. Many companies routinely run
geo-experiments to measure the effect of their marketing, generating
rich time-series data on outcomes and spending in both experimental
groups.

Still, it has not been clear how to best marry geo-experiments and MMMs
and how to efficiently utilize geo-experiments to calibrate
MMMs.\footnote{This problem exists in particular when multiple
  geo-experiments are available for the same marketing channel.} So far,
standard practice reduces each experiment to a single number, which is
then used to calibrate the MMM: total incremental revenue divided by
total incremental spend. This point estimate of Return on Advertising
Spend (ROAS) discards the temporal evolution of treatment
effects---precisely the variation needed to understand how marketing
effects persist over time (adstock) and how they diminish with scale
(saturation).

\textbf{Our contribution:} We show how to extract substantially more
information from geo-experimental time-series. Rather than reducing
experiments to point estimates, we directly use them to estimate the
structural parameters that govern marketing dynamics in MMM
specifications. Our approach recovers the complete functional form ---
adstock decay rates, saturation curves, and effectiveness coefficients
--- from the same time series of geo-experiments that standard analysis
reduces to a single ROAS number.

The key insight is simple: given the time series of outcomes for both
the treatment and control groups in geo-experiments, all confounding
factors cancel out when taking their difference. This leaves a clean
equation without any confounders that relates observed outcome
differences to observed marketing spends through exactly the
transformations MMM assumes. This differenced equation can be estimated
directly, yielding causally-identified estimates of all MMM parameters
governing the impact of marketing. Multiple geo-experiments can be
efficiently pooled to improve precision.

The remainder of this paper proceeds as follows. Section 2 discusses
related work and literature. Section 3 develops our key insight about
information in geo-experimental time-series and presents the structural
estimation approach. Section 4 presents the Bayesian estimation
framework. Section 5 demonstrates the approach on synthetic data.
Section 6 discusses integration with full MMM estimation and
limitations.

\section{Related Work}\label{related-work}

Our work engages three streams of literature: the identification
challenges in marketing mix modeling, experimental methods for
advertising measurement, and approaches to calibrating MMMs with
experimental data.

\textbf{Identification in Marketing Mix Models.} Modern MMMs capture
advertising dynamics through adstock transformations for temporal
carryover and saturation functions for diminishing returns, typically
estimated jointly in a Bayesian framework
(\citeproc{ref-jin2017bayesian}{Jin et al. 2017}). A fundamental
challenge, established by Dew et al. (\citeproc{ref-dew2024mmm}{2024}),
is that saturation effects and time-varying effectiveness are not
separately identifiable from observational data: models incorporating
either effect can approximate data generated by the other, yet imply
different optimal budget allocations. This identification failure is
particularly acute when spending follows autoregressive patterns---a
common feature of marketing data exacerbated by adstock transformations
themselves. Gordon et al. (\citeproc{ref-gordon2019comparison}{2019})
and Gordon et al. (\citeproc{ref-gordon2023close}{2023}) demonstrate
empirically that observational methods---including matching, regression
adjustment, and synthetic controls---fail to recover causal effects from
randomized experiments even with hundreds of millions of observations
and rich behavioral covariates. The implication is clear: experimental
variation, not merely richer data, is necessary for reliable
identification.

\textbf{Experimental Measurement Methods.} Geo-experiments enable
randomized treatment at the market level, avoiding the individual-level
tracking challenges that complicate user-based experiments. Vaver and
Koehler (\citeproc{ref-vaver2011geo}{2011}) introduce the geo-experiment
framework for measuring ad effectiveness through randomized regional
assignment. Brodersen et al.
(\citeproc{ref-brodersen2015causalimpact}{2015}) develop Bayesian
structural time-series models for counterfactual prediction, implemented
in the CausalImpact package. Chen and Au
(\citeproc{ref-chen2022trimmedmatch}{2022}) introduce the Trimmed Match
estimator for robust inference on incremental return on ad spend from
paired geo-experiments. The synthetic control framework of Abadie et al.
(\citeproc{ref-abadie2010synthetic}{2010}) provides foundations for
counterfactual construction when only aggregate data are available.
These methods yield credible estimates of total treatment effects, but
they do not recover the structural parameters---adstock decay,
saturation curvature, and channel effectiveness---that determine how
effects vary with spending levels and over time.

\textbf{Calibrating MMMs with Experiments.} Recognizing that
observational MMMs may produce unreliable estimates, practitioners have
developed calibration approaches using experimental results. Zhang et
al. (\citeproc{ref-zhang2024calibration}{2024}) reparameterize MMMs with
ROAS as a direct parameter, enabling informative Bayesian priors derived
from experiments. Meta's Robyn treats calibration as a third objective
in multi-objective optimization, penalizing deviation from experimental
lift estimates (\citeproc{ref-robyn2023}{Runge et al. 2023}). The
pymc-marketing package adds lift test observations to the model
likelihood, treating them as points on the saturation curve
(\citeproc{ref-orduz2024pymc}{Orduz 2024}). These approaches share a
common limitation: they reduce each experiment to an aggregate
statistic---a point estimate of total lift or ROAS---discarding the
temporal structure of experimental outcomes. The Bayesian prior approach
cannot distinguish experiments conducted at different spending levels
without ad-hoc weighting. The saturation-curve approach identifies the
shape of diminishing returns but not adstock dynamics, since it does not
use how effects evolve over time. In all cases, information is lost.

\section{Structural Estimation from
Geo-Experiments}\label{structural-estimation-from-geo-experiments}

\subsection{Geo-Experiments and Standard
Practice}\label{geo-experiments-and-standard-practice}

Geo-experiments divide a market into geographic units---typically
Designated Market Areas (DMAs) or metropolitan regions---and randomly
assign these units to a treatment or control group. During the test
period, marketing spend is modified in the treatment group (often
reduced to zero, in ``go-dark'' tests) while the control group maintains
normal spending.

A well-designed geo-experiment includes three phases:

\begin{itemize}
\tightlist
\item
  \textbf{Pre-test period}: Establishes that treatment and control
  groups follow parallel trends before intervention
\item
  \textbf{Test period}: Marketing differs between groups; outcome
  differences reflect causal effects
\item
  \textbf{Cooldown period}: Marketing returns to normal; residual
  effects decay over time
\end{itemize}

\begin{figure}
\centering
\pandocbounded{\includegraphics[keepaspectratio,alt={Time-series from a geo-experiment showing treatment and control group dynamics across test and cooldown periods. (A) Sales for treatment and control groups. (B) Marketing spend in both groups. (C) Sales difference (control minus treatment) highlighting the causal impact of advertising, with the true effect gap (dashed) for reference---available here because the data is synthetic. Shaded regions indicate the test period where marketing was turned off in the treatment group.}]{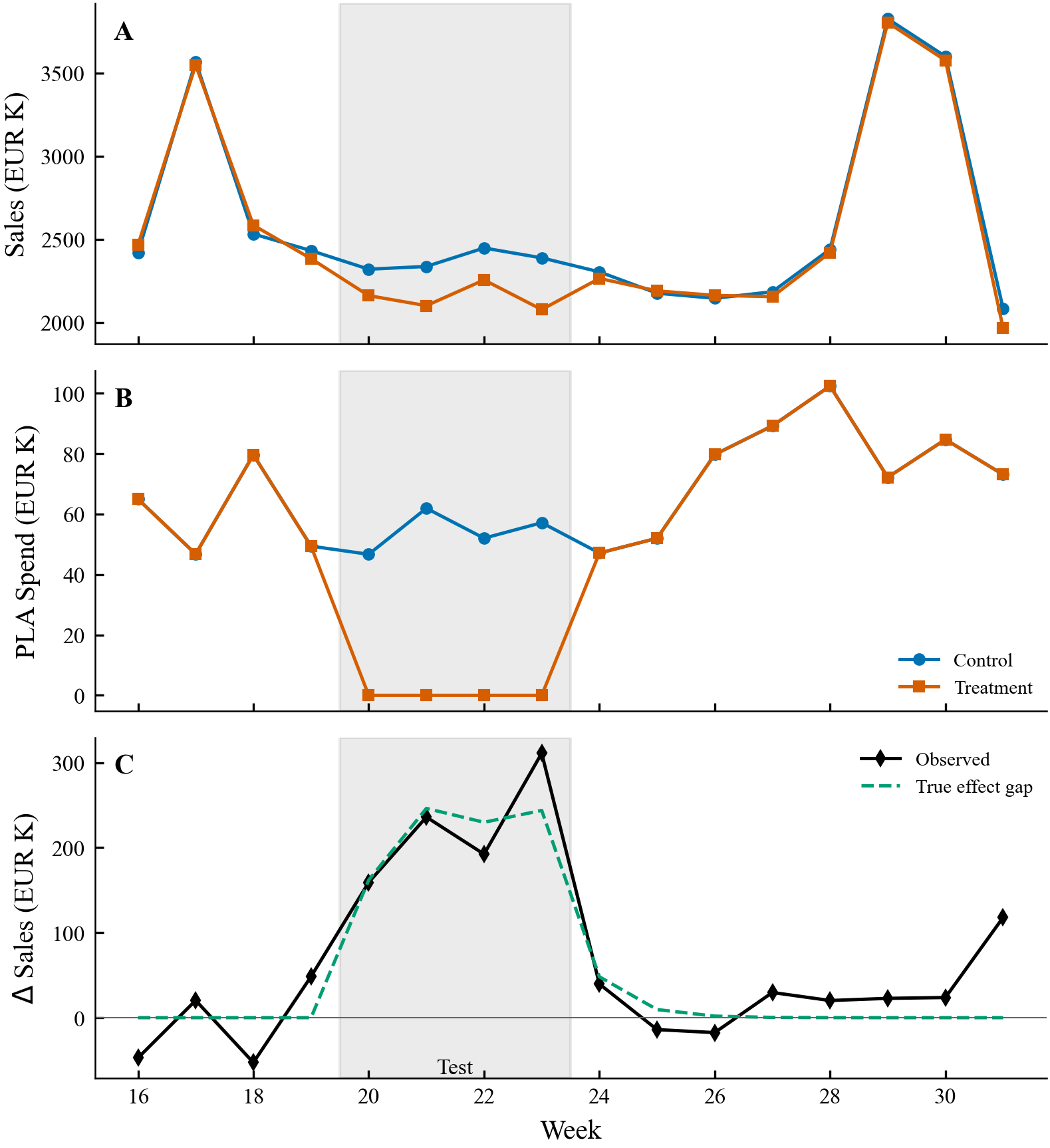}}
\caption{Time-series from a geo-experiment showing treatment and control
group dynamics across test and cooldown periods. (A) Sales for treatment
and control groups. (B) Marketing spend in both groups. (C) Sales
difference (control minus treatment) highlighting the causal impact of
advertising, with the true effect gap (dashed) for reference---available
here because the data is synthetic. Shaded regions indicate the test
period where marketing was turned off in the treatment
group.}\label{fig:geotest}
\end{figure}

The raw output of geo-experiments is rich: complete time-series of
outcomes and marketing spend for both groups, typically at daily
granularity. Yet standard practice computes a single number:

\[
\widehat{\text{ROAS}} = \frac{\sum_{t \in \text{test}} (y_t^{\text{control}} - y_t^{\text{treatment}})}{\sum_{t \in \text{test}} (x_t^{\text{control}} - x_t^{\text{treatment}})}
\]

This point estimate answers the question: ``What was the average return
on this marketing spend?'' But by summing over the test period and
dividing totals, it discards all information about \emph{dynamics}---how
effects evolve over time and how they respond to different spending
levels.

\subsection{Marketing Dynamics in Theory and in Geo-Test
Data}\label{marketing-dynamics-in-theory-and-in-geo-test-data}

Marketers have long recognized that advertising effects are neither
immediate nor constant. Effects \emph{persist}: a TV ad seen this week
may influence a purchase next week. And there are \emph{saturation
effects}: the hundredth impression on the same audience yields less
incremental impact than the first.\footnote{The notion of diminishing
  returns is essential to the idea of marketing mix modeling: if returns
  were constant across channels, no mix would be required---one would
  simply allocate the entire budget to the highest-return channel.}

Marketing Mix Models formalize these dynamics by applying two
transformations to translate marketing spends into incremental sales:

\textbf{Adstock} captures persistence and delayed effects: the impact of
marketing in week \(t\) is determined not only by the marketing spend of
week \(t\), but also the marketing spends in previous weeks. The first
transformation is hence to calculate the adstock - which can be thought
of as measuring how present the ads are in the mind of consumers in week
\(t\) - as a function of the marketing spends in the current and the
\(L-1\) previous weeks.

\[
x^*_t = \frac{\sum_{\ell=0}^{L-1} \alpha^\ell \, x_{t-\ell}}{\sum_{\ell=0}^{L-1} \alpha^\ell}
\]

Here \(x_{t-\ell}\) is marketing spending \(\ell\) weeks ago, \(L\) is
the lookback window (e.g., 6 weeks), and \(\alpha \in [0,1)\) is the
decay rate, indicating how much spending from previous weeks `carries
over.' When \(\alpha = 0\), only current-week spending matters; when
\(\alpha = 0.7\), spending from several weeks ago still contributes
substantially.\footnote{In other words, the adstock \(x^*_t\) in a given
  week \(t\) is determined by the marketing spend in week \(t\) and the
  \(L-1\) previous weeks. As can be seen in the equation, each past
  week's spending is weighted by \(\alpha^\ell\), so more recent
  spending counts more. The denominator normalizes the weights to sum to
  one, ensuring that €1000 of spending produces €1000 of total adstock,
  distributed across weeks rather than concentrated in one.}

\textbf{Saturation} captures diminishing returns to adstock. Combined
with an effectiveness coefficient \(\beta\), the saturation function
translates the adstock in week \(t\) into the impact of marketing in
week \(t\):

\[
IncrementalSales_t = \beta \cdot \text{sat}(x^*_t; \lambda) 
\]

\(\text{sat}(x^*_t; \lambda)\) is generally chosen to be an increasing,
concave function, such as the logistic saturation function, which rises
steeply at low adstock and flattens at high adstock.\footnote{We use the
  logistic saturation function,
  \(\text{sat}(x^*_t; \lambda) =  \frac{1 - e^{-\lambda x^*_t}}{1 + e^{-\lambda x^*_t}}\)}
The parameter \(\lambda > 0\) controls the curvature: higher \(\lambda\)
means saturation (diminishing returns) set in earlier, at lower levels
of adstock. The coefficient \(\beta\) then converts the saturated
adstock into euros of incremental revenue.

Three parameters thus govern how spend on a given marketing channel
translates into incremental sales: \(\alpha\) (persistence of spending),
\(\lambda\) (speed of diminishing returns), and \(\beta\) (overall
effectiveness).

\begin{figure}
\centering
\pandocbounded{\includegraphics[keepaspectratio,alt={The two key transformations in marketing mix models. (A) Adstock decay weights for different \textbackslash alpha values, showing how past spending contributes to current-period impact. (B) Saturation curves for different \textbackslash lambda values, showing diminishing returns to adstock.}]{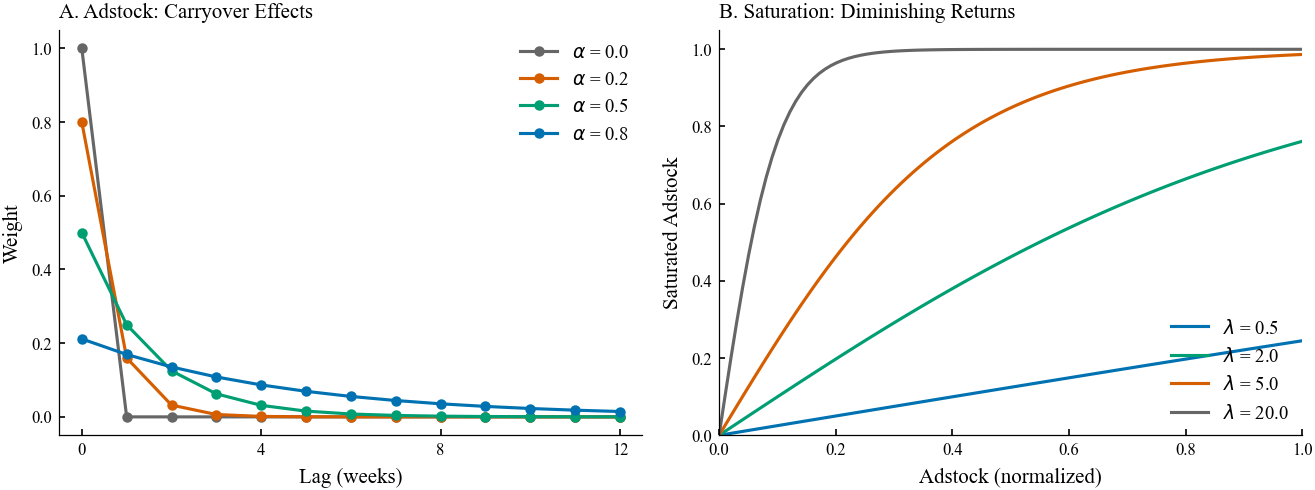}}
\caption{The two key transformations in marketing mix models. (A)
Adstock decay weights for different \(\alpha\) values, showing how past
spending contributes to current-period impact. (B) Saturation curves for
different \(\lambda\) values, showing diminishing returns to
adstock.}\label{fig:mmm-transformations}
\end{figure}

\textbf{What geo-tests reveal.} Returning to fig.~\ref{fig:geotest},
note that time-series data from geo-experiments reveals these dynamics
directly.

\emph{Adstock from the test and cooldown periods.} The adstock mechanism
is visible at both ends of the experiment. When marketing stops in the
treatment group at the start of the go-dark test period, the outcome gap
doesn't emerge instantly---it widens gradually as treatment's adstock
depletes while control's is maintained. Similarly, when marketing
resumes at the start of the cooldown period, the gap doesn't close
instantly---it narrows gradually as treatment's adstock rebuilds. The
rate of these transitions identifies \(\alpha\). Fast
transitions---where the gap emerges or closes within a week or
two---imply low \(\alpha\): spending in previous weeks carries over very
little, so adstock depletes and rebuilds quickly. Slow
transitions---where convergence takes many weeks---imply high
\(\alpha\): past spending continues to influence outcomes for an
extended period, so adstock changes gradually.

\emph{Saturation from spending variation.} Geo-tests at different
spending levels trace the saturation curve. If doubling spend doubles
the outcome gap, returns are proportional (\(\lambda\) is small). If
doubling spend yields less than double the gap, we observe diminishing
returns (\(\lambda\) is larger).

Standard ROAS analysis discards both signals entirely. Our approach
recovers them.

\subsection{The Differencing Equation}\label{the-differencing-equation}

We recover this information by estimating the structural model directly
on geo-test data. The key is recognizing that treatment and control
groups can be viewed as two parallel universes: identical in every
respect except for marketing in the tested channel.

\textbf{The parallel universes.} Consider a geo-experiment testing
marketing channel \(A\). If our specification correctly describes how
marketing spends translate into incremental sales, outcomes in each
group follow:

\[
y_t^{\text{control}} = \mu_t + \beta_A \cdot \text{sat}(x^{*,\text{control}}_{t,A}; \lambda_A) + \varepsilon_t^{\text{control}}
\]

\[
y_t^{\text{treatment}} = \mu_t + \beta_A \cdot \text{sat}(x^{*,\text{treatment}}_{t,A}; \lambda_A) + \varepsilon_t^{\text{treatment}}
\]

Here \(\mu_t\) captures everything except marketing spends on channel
\(A\): baseline demand, contributions from other marketing channels,
effects of pricing and promotions, seasonality, and any unobserved
factors.\footnote{The full MMM specification models outcomes as:
  \[y_t = \beta_0 + \sum_{m=1}^{M} \beta_m \cdot \text{sat}(x^*_{t,m}; \lambda_m) + \sum_{c} \gamma_c z_{t,c} + \sum_{u} \delta_u w_{t,u} + \varepsilon_t\]
  where the sum over \(m\) covers all \(M\) marketing channels,
  \(z_{t,c}\) are observed controls (seasonality, pricing, etc.),
  \(w_{t,u}\) are unobserved factors affecting demand (competitor
  actions, macroeconomic sentiment, etc.), and \(\varepsilon_t\) is
  idiosyncratic noise. When we isolate channel \(A\), we can rewrite
  this as
  \(y_t = \mu_t + \beta_A \cdot \text{sat}(x^*_{t,A}; \lambda_A) + \varepsilon_t\),
  where:
  \[\mu_t = \beta_0 + \sum_{m \neq A} \beta_m \cdot \text{sat}(x^*_{t,m}; \lambda_m) + \sum_{c} \gamma_c z_{t,c} + \sum_{u} \delta_u w_{t,u}\]
  That is, \(\mu_t\) contains everything except channel \(A\)'s
  contribution---including the unobserved factors \(w_{t,u}\). Because
  randomization balances these unobserved factors across treatment and
  control groups, they cancel when we difference, even though we never
  observe them directly. See Appendix A for a formal treatment of
  identification.} Crucially, \(\mu_t\) is \emph{identical} in both
groups because randomization ensures treatment and control differ only
in channel \(A\) spending.

\textbf{The differencing equation.} Subtracting treatment from control
eliminates \(\mu_t\):

\[
\begin{aligned}
y_t^{\text{control}} - y_t^{\text{treatment}} &= \beta_A \bigl[ \text{sat}(x^{*,\text{control}}_{t,A}; \lambda_A)  \\
&\quad 
- \text{sat}(x^{*,\text{treatment}}_{t,A}; \lambda_A) \bigr] + \nu_t
\end{aligned}
\]

where
\(\nu_t = \varepsilon_t^{\text{control}} - \varepsilon_t^{\text{treatment}}\).

The intercept cancels; other channels cancel (as spends are identical in
both groups); control variables cancel (as values are identical in both
groups); unobserved confounders cancel (as, balanced by randomization,
they are identical in both groups). What remains is a function of only
the channel \(A\) parameters---\(\alpha_A\) (embedded in the adstock
\(x^*\)), \(\lambda_A\), and \(\beta_A\)---and observed spending data.

\textbf{The key property.} The differencing equation contains only three
unknown parameters and only observable quantities: the outcome
difference on the left, spending in both groups on the right. Everything
else---seasonality, other channels, unobserved demand shocks---has
vanished through differencing.

This resolves the identification problem that plagues observational MMM
estimation. To estimate causal effects from observational data in a
standard MMM, we must account for all factors that influence both
marketing spending and outcomes. For example, if high expected demand
causes both higher sales and higher marketing budgets (because firms
advertise more when they anticipate strong demand), we must observe and
control for ``expected demand'' perfectly---otherwise, we cannot
distinguish whether sales increased because of marketing or because
demand was high. In econometric terms, any factor that affects both
spending and outcomes but is not controlled for ends up in the error
term, violating the assumption that errors are independent of spending.

In practice, this assumption is nearly impossible to satisfy. Marketing
budgets respond to expected demand, algorithmic bidding systems chase
performance signals, and strategic decisions depend on factors that
researchers cannot fully observe. The differencing estimator sidesteps
this problem entirely: randomization ensures that all
confounders---observed and unobserved---are balanced across treatment
and control groups, so they cancel through differencing. We achieve
causal identification through experimental design rather than by hoping
our controls capture every relevant factor.

\section{Estimation}\label{estimation}

The differencing equation provides the foundation for estimation. We
observe outcome differences and spending in both groups; we seek to
recover the parameters (\(\alpha\), \(\lambda\), \(\beta\)) that govern
the channel's effect. This section describes how to prepare data for
estimation, how to pool observations across multiple geo-experiments,
and our Bayesian estimation approach.

\subsection{Data Preparation}\label{data-preparation}

\subsubsection{From the Differencing Equation to Observable
Data}\label{from-the-differencing-equation-to-observable-data}

The differencing equation makes clear what data we need for each
observation. The left-hand side requires outcomes in both treatment and
control groups at time \(t\), so we can compute their difference. The
right-hand side requires marketing spend in both groups---not just at
time \(t\), but also in the preceding \(L-1\) weeks, since these
determine the adstock \(x^*_t\).

For each time period \(t\), we therefore need:

\begin{itemize}
\tightlist
\item
  \(y_t^{\text{control}}\) and \(y_t^{\text{treatment}}\) (to compute
  the outcome difference)
\item
  \(x_t^{\text{control}}, x_{t-1}^{\text{control}}, \ldots, x_{t-L+1}^{\text{control}}\)
  (to compute control adstock)
\item
  \(x_t^{\text{treatment}}, x_{t-1}^{\text{treatment}}, \ldots, x_{t-L+1}^{\text{treatment}}\)
  (to compute treatment adstock)
\end{itemize}

\subsubsection{Wide Format}\label{wide-format}

We reshape the geo-test data into \emph{wide format}, where each row
contains all the information needed for one observation: the outcome
difference and the current plus lagged spending for both groups.

This format has a crucial property: each row is now a self-contained
observation with no temporal dependencies on other rows. All the lagged
spending values needed to compute adstock are included within the row
itself.

\subsubsection{Pooling Multiple
Experiments}\label{pooling-multiple-experiments}

The wide format makes pooling across multiple geo-experiments
straightforward: we simply stack the rows from different tests. Since
each row contains all information needed for estimation (including
lagged spends), there are no cross-row dependencies to worry about.
Observations from a geo-test run in January can be combined with
observations from a test run in June---they are simply additional rows
in the dataset, each providing independent information about the same
underlying parameters (\(\alpha\), \(\lambda\), \(\beta\)).

\subsubsection{Scaling}\label{scaling}

The parameters we estimate are not scale-invariant. A geo-test run on a
subset of the country will have smaller absolute spending and outcomes
than the country-level MMM we ultimately want to calibrate. To ensure
our parameter estimates are comparable, we transform the geo-test data
to match the scale of the MMM:

\begin{enumerate}
\def\labelenumi{\arabic{enumi}.}
\item
  \textbf{Scale to country level.} If the treatment group represents
  40\% of the country and the control group 60\%, we divide treatment
  outcomes and spending by 0.4 and control outcomes and spending by 0.6.
  This transforms the data to what we would observe if each group were
  the entire country.
\item
  \textbf{Apply MMM scaling.} In case the MMM applies data
  transformations for numerical stability (e.g., dividing outcomes
  and/or spends by their maximum value), we apply the same
  transformation to the geo-test data. This ensures parameters estimated
  from geo-tests are on the same scale as those of the MMM.
\end{enumerate}

\subsection{Bayesian Model}\label{bayesian-model}

We estimate the differencing equation using Bayesian inference with
weakly informative priors:

\begin{table}[!htbp]
\centering
\footnotesize
\caption{Bayesian model priors}\label{tbl-priors}
\begin{tabular}{@{}lll@{}}
\toprule
Parameter & Prior & Rationale \\
\midrule
$\alpha$ & Beta(1, 3) & Limited carryover for digital channels \\
$\lambda$ & Gamma(3, 1) & Positive, mode at 2 \\
$\beta$ & HalfNormal(2) & Positive effectiveness \\
$\sigma$ & HalfNormal(0.5) & Observation noise \\
\bottomrule
\end{tabular}
\end{table}

These are the default priors of the \texttt{pymc-marketing} package
(v0.19). Deliberately, the observational MMM benchmark of Section 5 uses
the same priors for its media parameters, so the comparison between the
two methods cannot be confounded by prior choice; Appendix C examines
sensitivity to these choices for both models. The likelihood follows
directly from the differencing equation: the observed outcome difference
equals the predicted difference in saturated adstock effects plus
normally distributed noise.

Implementation details, including the full probabilistic model in plate
notation, are provided in Appendix B.

\section{Empirical Demonstration}\label{empirical-demonstration}

\subsection{Synthetic Data}\label{synthetic-data}

We demonstrate the approach using synthetic data designed to capture
realistic features of marketing measurement. The data simulates an
online retailer with three marketing channels (paid search, social
media, TV) over 156 weeks, incorporating:

\begin{itemize}
\tightlist
\item
  Realistic seasonality and trends
\item
  Endogenous marketing allocation (budgets respond to expected demand)
\item
  Algorithmic bidding that chases performance signals
\item
  True saturation and adstock effects with known parameters
\end{itemize}

The synthetic data generation process is described in detail in a
companion paper (\citeproc{ref-heusch2026synthetic}{Heusch 2026}), whose
notebook generates the exact dataset used here. All data is synthetic:
parameter values, magnitudes, and business mechanisms are fictional,
chosen to lie in the range of published industry benchmarks, and the
complete data-generating process---including the true value of every
parameter---is public. The key advantage is that we observe the true
data-generating process, including ground-truth values for all
parameters.

The true parameters for the paid search (PLA) channel are: adstock decay
\(\alpha = 0.2\) (20\% weekly carryover); saturation \(\lambda = 0.008\)
per thousand euros of adstocked spend, implying half-saturation at
roughly EUR 137K of weekly adstocked spend; effectiveness
\(\beta = 1{,}100\) (thousands of euros per unit of saturated adstock);
and \textbf{true ROAS of 4.20×}. Because estimation divides each series
by its maximum (Section 4), estimated parameters live on that scale; the
scaled truths are \(\lambda_{\text{scaled}} \approx 1.71\) and
\(\beta_{\text{scaled}} \approx 0.18\), and the parameter tables below
report this scale.

\subsection{Simulated Geo-Tests}\label{simulated-geo-tests}

From the synthetic data, we generate four geo-experiments for the PLA
channel at different time periods and spending levels:

\begin{enumerate}
\def\labelenumi{\arabic{enumi}.}
\tightlist
\item
  \textbf{Test 1} (weeks 20--23): low spending (mean control-group PLA
  spend of EUR 54K per week)
\item
  \textbf{Test 2} (weeks 55--58): high spending (EUR 129K)
\item
  \textbf{Test 3} (weeks 100--103): moderate spending (EUR 64K)
\item
  \textbf{Test 4} (weeks 140--143): low-to-moderate spending (EUR 57K)
\end{enumerate}

Each test includes 4 weeks pre-test, 4 weeks test (with PLA spend set to
zero in treatment), and 8 weeks cooldown. Estimation uses all 16 weeks
of every test---pre-period, test, and cooldown---with lagged spends
taken from the full spend history, so the gap-opening and gap-closing
transitions enter the likelihood. The variation in spending levels
across tests aids identification of the saturation parameter: jointly,
the four tests probe adstocked spending from roughly EUR 43K to EUR 194K
per week, nearly the full range observed over the three years. Both
experimental groups are simulated directly at country scale (as if each
covered the entire market), which spares the rescaling step of Section 4
without changing anything else.

fig.~\ref{fig:geotest} shows the time-series for Test 1. During the test
period, treatment group sales fall relative to control, reflecting the
causal effect of PLA advertising. The gap closes almost immediately in
the cooldown weeks---consistent with the low adstock effect (\(\alpha\)
= 0.2).

\subsection{Main Results: ROAS
Recovery}\label{main-results-roas-recovery}

We now evaluate whether our structural estimation approach can recover
the true ROAS---the ground truth parameter known only in
simulation---from the experimentally observed treatment effects.

\begin{table}[!htbp]
\centering
\footnotesize
\caption{ROAS Estimates Comparison.}\label{tbl-roas2}
\begin{tabular}{@{}lrcc@{}}
\toprule
Method & ROAS & 90\% CI & In CI? \\
\midrule
\textbf{True ROAS} & 4.20× & — & — \\
MMM (realistic) & 10.66× & [6.99, 14.42] & No \\
MMM (oracle ctrl.) & 8.45× & [6.97, 9.82] & No \\
Structural (2 tests) & 4.31× & [3.87, 4.74] & Yes \\
Structural (4 tests) & 4.14× & [3.77, 4.51] & Yes \\
\bottomrule
\end{tabular}
\par\vspace{0.8em}
\begin{minipage}{\columnwidth}
\footnotesize
\textit{Note:} ``MMM (realistic)'' is the observational specification a practitioner would run: observed promotion dummy, observed promotional price level, and three annual Fourier harmonics for seasonality—the standard seasonal controls in Robyn, Meridian, and pymc-marketing. ``MMM (oracle ctrl.)'' replaces these with the true confounders from the data-generating process (true promotion state, true price level, seasonal component, product-quality drift, market sentiment)—covariates no practitioner possesses, providing an upper bound on what better controls could achieve. The `In CI?' column indicates whether the true ROAS (4.20×) falls within the 90\% credible interval.
\end{minipage}
\end{table}

\begin{figure}
\centering
\pandocbounded{\includegraphics[keepaspectratio,alt={PLA ROAS estimates with 90\% credible intervals. Ground truth (green), observational MMM with realistic and with oracle controls (pink, gray), and structural estimation with 2 and 4 geo-tests (blue, orange). Both MMM specifications overestimate ROAS due to endogenous spend---the oracle showing that even perfect controls cannot repair it; structural estimation recovers the true value.}]{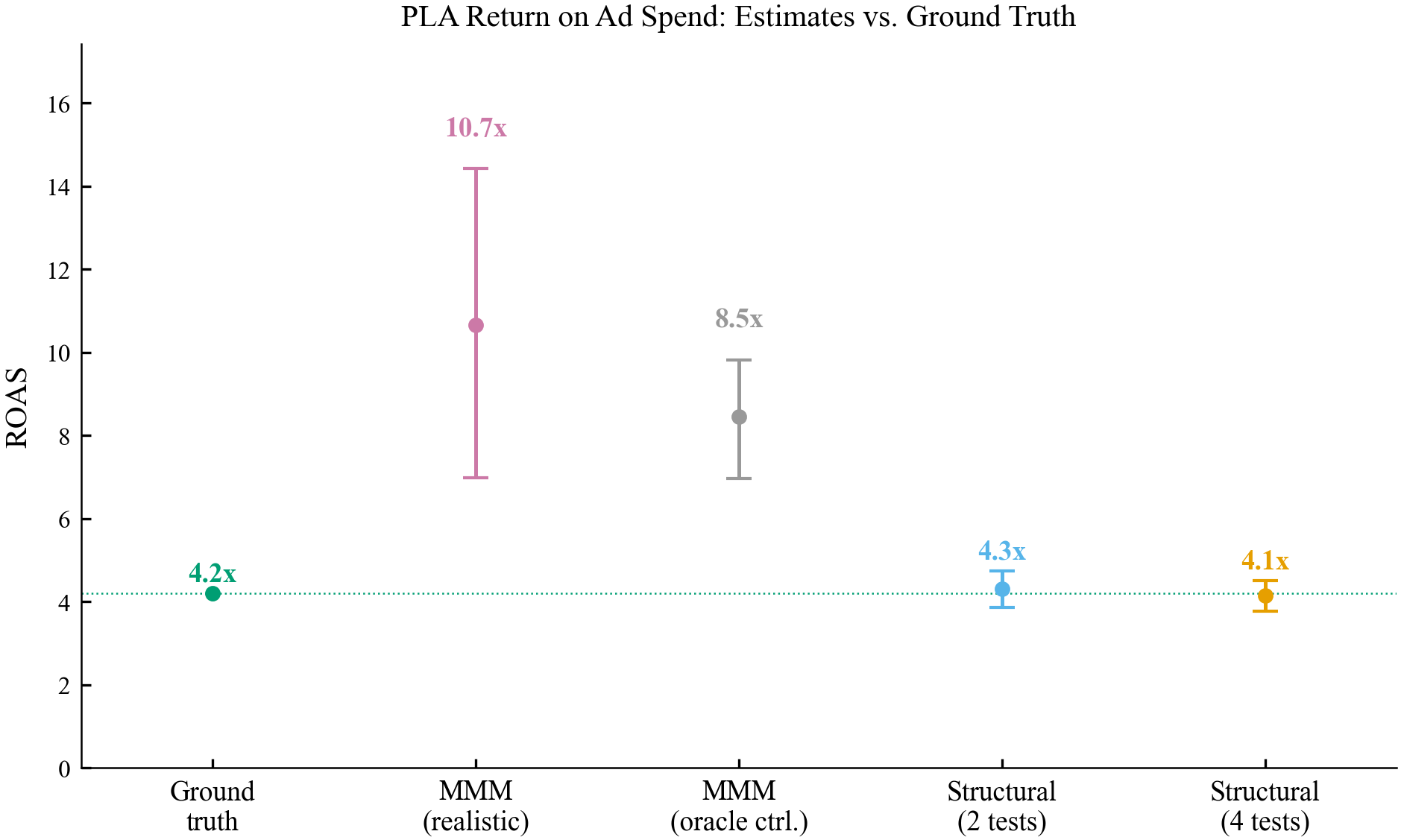}}
\caption{PLA ROAS estimates with 90\% credible intervals. Ground truth
(green), observational MMM with realistic and with oracle controls
(pink, gray), and structural estimation with 2 and 4 geo-tests (blue,
orange). Both MMM specifications overestimate ROAS due to endogenous
spend---the oracle showing that even perfect controls cannot repair it;
structural estimation recovers the true
value.}\label{fig:roas-comparision}
\end{figure}

As observational benchmarks, we estimate the MMM of Section 3 on the
full three years of data in two specifications that differ only in their
information set. \emph{MMM (realistic)} conditions on what a
practitioner actually observes: the observed promotion dummy, the
observed promotional price level, and three annual Fourier harmonics for
seasonality---the standard seasonal device in Robyn, Meridian, and
pymc-marketing. \emph{MMM (oracle controls)} conditions instead on the
true confounders of the data-generating process---the true promotion
state and price level, and the latent seasonal, product-quality, and
market-sentiment components---covariates no practitioner could possess.
The oracle is not a feasible model but a diagnostic: it bounds what any
improvement in controls could achieve, so any bias that survives it
cannot be an omitted-variables problem.

The realistic MMM produces a severely biased ROAS estimate (10.66×
vs.~true 4.20×)---overestimating effectiveness by a factor of roughly
2.5---despite carrying the seasonal and promotional controls of standard
practice. The cause is endogeneity in the synthetic data: marketing
spend correlates with demand through strategic planning and algorithmic
optimization, and the model attributes the correlation to causation. The
oracle specification shows that this cannot be repaired by better
controls. Handed the true confounders themselves, the model still
estimates 8.45×---about double the true return, with an interval that
excludes it. Two mechanisms survive even perfect controls: the bidding
algorithm responds to \emph{realized} weekly performance, including its
random component, which no covariate an analyst could hold constant can
absorb; and the true baseline combines its components multiplicatively
while the controls enter linearly, leaving residual variation that the
endogenous spend tracks. The structural estimation approach recovers
ROAS close to the true value (4.31× with 2 tests, 4.14× with 4 tests),
with the true ROAS falling within the 90\% credible interval in both
cases. Precision improves as more geo-tests are pooled, but even two
experiments provide accurate point estimates.

\subsection{Parameter Estimates}\label{parameter-estimates}

Beyond ROAS, the structural approach recovers the underlying MMM
parameters:

\begin{table}[!htbp]
\centering
\footnotesize
\caption{Parameter estimates comparison}\label{tbl-params}
\begin{tabular}{@{}lccc@{}}
\toprule
Method & $\alpha$ (adstock) & $\lambda$ (saturation) & $\beta$ (effectiveness) \\
\midrule
\textbf{True} & 0.20 & 1.71 & 0.18 \\
Structural (2 tests) & 0.18 [0.08, 0.29] & 2.28 [0.91, 3.48] & 0.17 [0.11, 0.32] \\
Structural (4 tests) & 0.19 [0.09, 0.28] & 1.94 [0.65, 3.10] & 0.20 [0.11, 0.44] \\
MMM (realistic) & 0.50 [0.37, 0.62] & 1.27 [0.38, 2.62] & 0.86 [0.30, 2.03] \\
\bottomrule
\end{tabular}
\par\vspace{0.8em}
\begin{minipage}{\columnwidth}
\footnotesize
\textit{Note:} Values in brackets are 90\% credible intervals. Parameters are reported on the estimation scale (Section 4): $\alpha$ is scale-free; $\lambda$ is multiplied by the maximum weekly PLA spend (EUR 214K); $\beta$ is divided by the maximum weekly sales (EUR 6.03M).
\end{minipage}
\end{table}

\begin{figure}
\centering
\pandocbounded{\includegraphics[keepaspectratio,alt={Posterior distributions for \textbackslash alpha, \textbackslash lambda, and \textbackslash beta (on the estimation scale). True values indicated by vertical lines. Structural estimation (blue, orange) produces posteriors covering the ground truth; the observational MMM (pink) shows substantial bias, particularly for the adstock and effectiveness parameters.}]{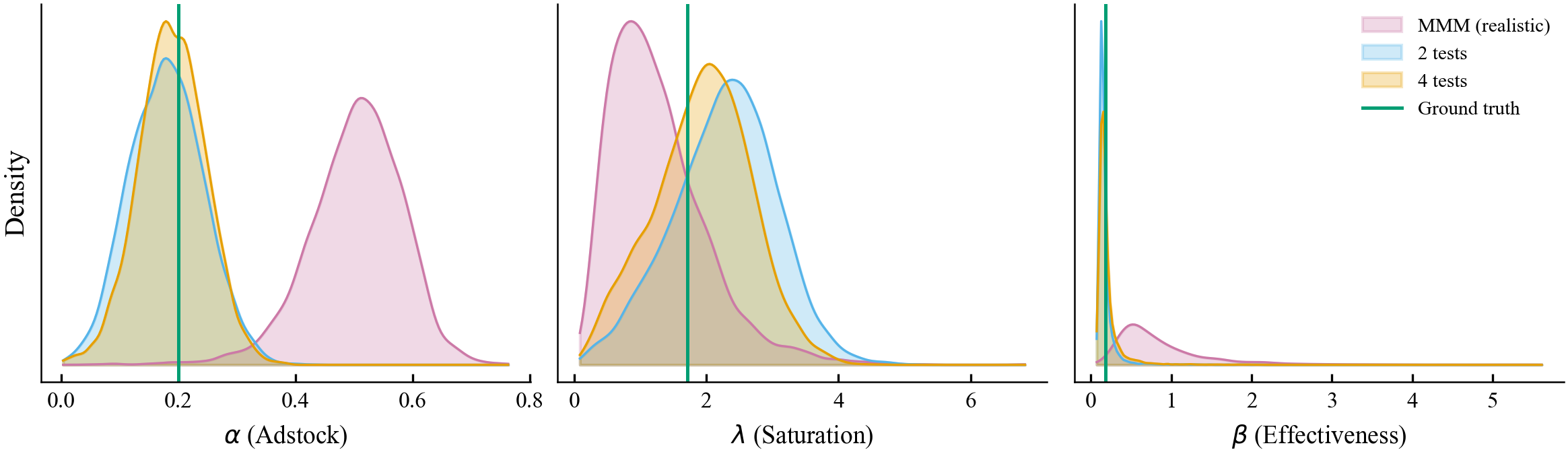}}
\caption{Posterior distributions for \(\alpha\), \(\lambda\), and
\(\beta\) (on the estimation scale). True values indicated by vertical
lines. Structural estimation (blue, orange) produces posteriors covering
the ground truth; the observational MMM (pink) shows substantial bias,
particularly for the adstock and effectiveness
parameters.}\label{fig:posteriors}
\end{figure}

\begin{figure}
\centering
\pandocbounded{\includegraphics[keepaspectratio,alt={Response curve identified by the experiments. Posterior draws and posterior mean (orange) from the 4-test structural estimation against the true curve (green, dashed). Tick marks show the adstocked spending levels the experiments probed. Over that range the curve is tightly identified; beyond it, the draws fan out---an honest statement of what the experiments do and do not pin down.}]{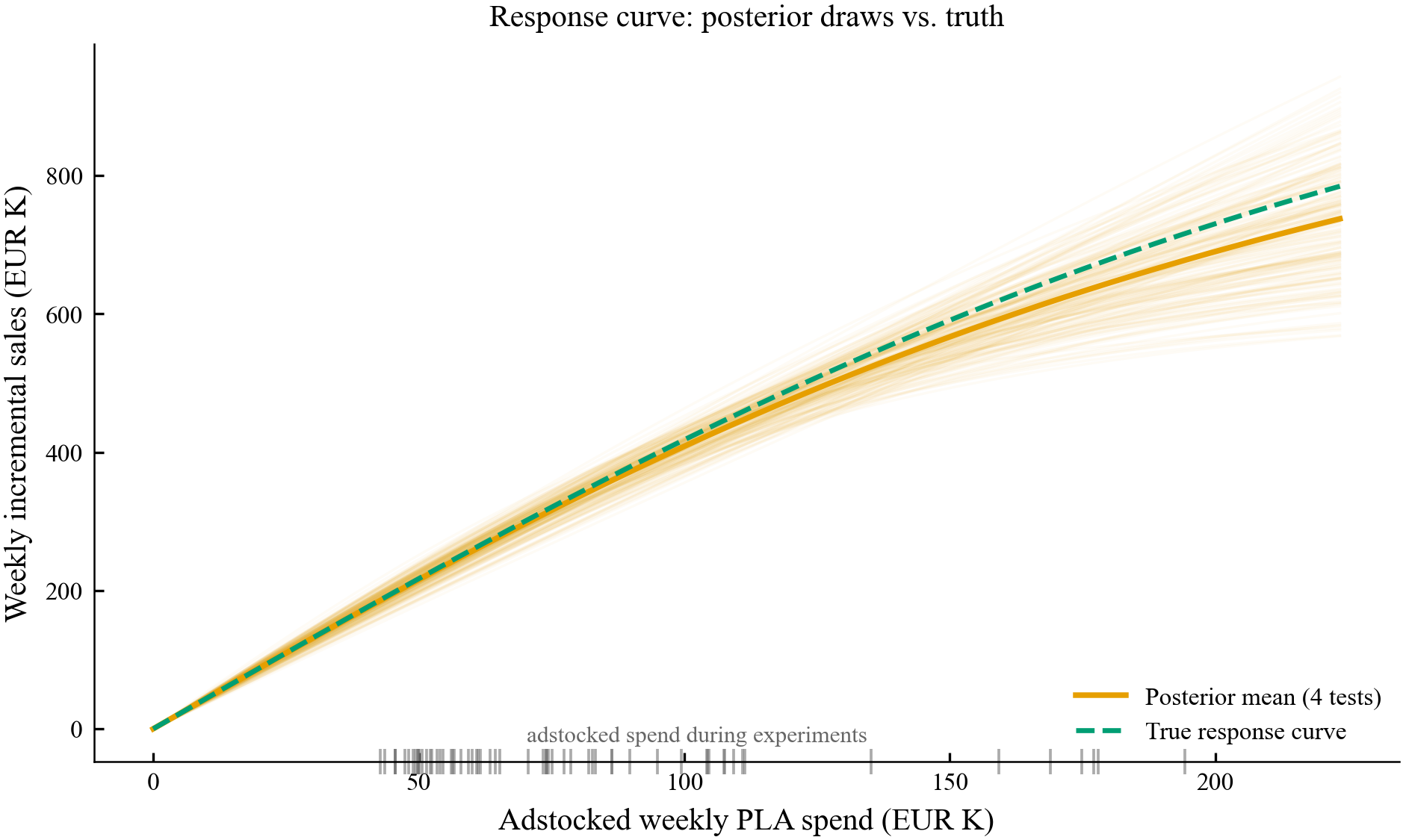}}
\caption{Response curve identified by the experiments. Posterior draws
and posterior mean (orange) from the 4-test structural estimation
against the true curve (green, dashed). Tick marks show the adstocked
spending levels the experiments probed. Over that range the curve is
tightly identified; beyond it, the draws fan out---an honest statement
of what the experiments do and do not pin down.}\label{fig:response}
\end{figure}

\begin{figure}
\centering
\pandocbounded{\includegraphics[keepaspectratio,alt={Pairwise joint posteriors from the 4-test structural estimation, with the true parameter values marked (×). The \textbackslash alpha panels are compact: the adstock rate is identified separately, by the gap-opening and gap-closing transitions. The \textbackslash lambda--\textbackslash beta panel shows the identification ridge: flat-but-large and steep-but-small parameter combinations imply nearly the same response over the tested spending range, so the data constrain the pair jointly rather than each alone---the ridge's invariant is the response curve of fig.~, and the ROAS it implies.}]{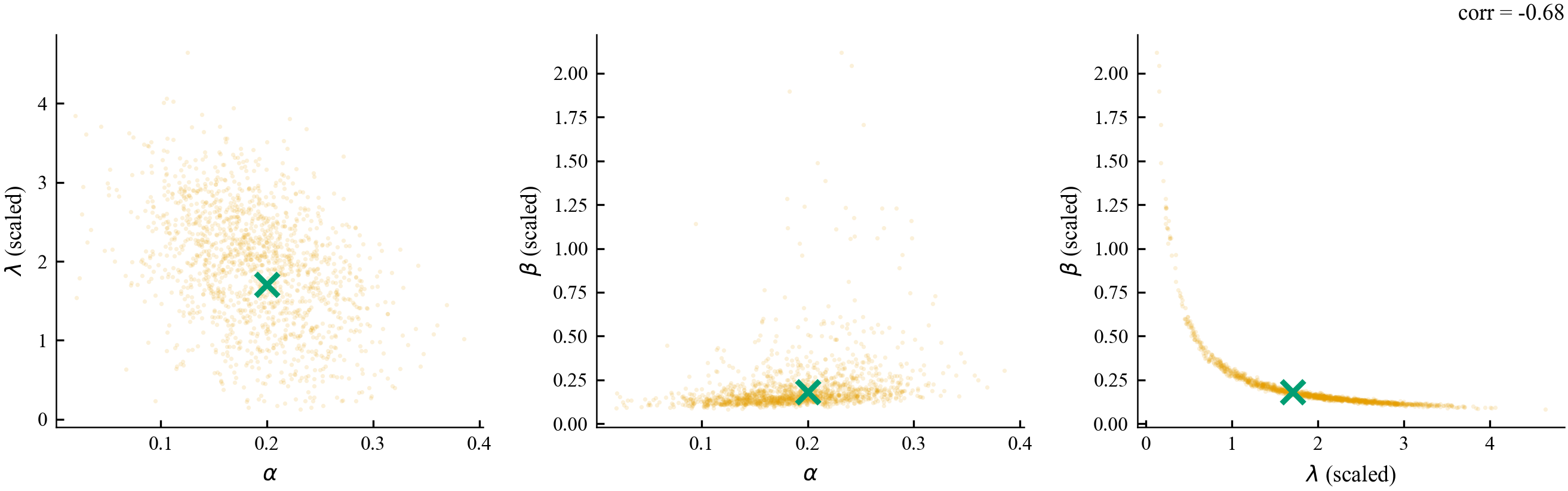}}
\caption{Pairwise joint posteriors from the 4-test structural
estimation, with the true parameter values marked (×). The \(\alpha\)
panels are compact: the adstock rate is identified separately, by the
gap-opening and gap-closing transitions. The \(\lambda\)--\(\beta\)
panel shows the identification ridge: flat-but-large and steep-but-small
parameter combinations imply nearly the same response over the tested
spending range, so the data constrain the pair jointly rather than each
alone---the ridge's invariant is the response curve of
fig.~\ref{fig:response}, and the ROAS it implies.}\label{fig:joint}
\end{figure}

The structural estimation recovers all three parameters: the true values
fall inside every 90\% credible interval, and the adstock posterior
centres almost exactly on the truth (\(\alpha\) = 0.19 against
0.20)---the information lives in the pre-period and transition weeks of
each test, which enter the likelihood in full. The saturation and
effectiveness posteriors are individually wider, for a structural
reason: \(\lambda\) and \(\beta\) trade off along a ridge
(fig.~\ref{fig:joint}, right panel; posterior correlation -0.68),
because a flatter curve with a larger coefficient produces nearly the
same response as a steeper curve with a smaller one over the range of
spending the experiments cover. What the data pin down tightly is the
ridge's joint implication---the response curve over the tested range,
and the ROAS it implies---which is why the ROAS intervals of Section 5.3
are far narrower than the marginal parameter intervals.
fig.~\ref{fig:response} shows this directly: over the tested range, the
posterior response curves lie on top of the truth, and they fan out only
beyond it.

In contrast, the realistic MMM produces substantially biased estimates,
particularly for the adstock parameter (0.50 vs.~true 0.20) and the
effectiveness coefficient (several times the truth). These parameter
biases combine to produce the inflated ROAS estimate.

\section{Discussion}\label{discussion}

Our simulation results demonstrate that the structural estimation
approach recovers all three parameters governing marketing
dynamics---adstock decay, saturation, and effectiveness---from
geo-experimental time-series data. This section situates our
contribution within the broader literature, discusses integration with
full MMM workflows, and acknowledges limitations.

\subsection{Relationship to the Experimental Measurement
Literature}\label{relationship-to-the-experimental-measurement-literature}

Our empirical results corroborate the finding of Gordon et al.
(\citeproc{ref-gordon2019comparison}{2019}) that observational methods
fail to recover causal advertising effects. Section 5.3 demonstrates
this directly: the observational MMM produces a severely biased ROAS
estimate despite carrying the standard seasonal and promotional
controls, and the oracle specification---handed the true confounders
themselves---reduces the bias only from 10.66× to 8.45× against a true
4.20×. When even perfect controls cannot restore identification, the
failure is not a matter of data quality. As we formalize in Appendix
A.2, the conditional independence assumption required for causal
identification---that error terms be independent of marketing spending
conditional on controls---is virtually impossible to satisfy when
budgets respond to anticipated demand, algorithmic systems chase
performance signals, and strategic decisions depend on unobservable
factors.

Our contribution builds on this foundation. We do not merely accept that
experiments are necessary; we demonstrate why observational MMM fails
and then show how to extract substantially more information from
experimental data than current practice allows. Rather than reducing
geo-experiments to aggregate ROAS estimates, we use the full time-series
to identify all structural parameters governing marketing dynamics. This
represents a shift from ``experiments for measurement'' to ``experiments
for structural estimation.''

The practical implication is considerable. Many firms already run
geo-experiments but discard most of the information they generate. Our
approach enables these firms to calibrate complete MMM functional
forms---adstock decay, saturation curvature, and effectiveness---from
the same experimental investments they already make.

\subsection{Comparison to Existing Calibration
Methods}\label{comparison-to-existing-calibration-methods}

Current MMM software packages offer various approaches to incorporating
experimental results, but all share a fundamental limitation: they
reduce geo-experiments to aggregate statistics, discarding the temporal
structure that identifies dynamic parameters.

\textbf{Google Meridian} (\citeproc{ref-zhang2024calibration}{Zhang et
al. 2024}) reparameterizes the model to include ROAS as a direct
parameter, enabling experimental point estimates to serve as informative
priors. This elegantly incorporates aggregate experimental evidence into
Bayesian estimation, but cannot distinguish between experiments
conducted at different spending levels or time horizons without ad-hoc
weighting. The approach identifies the level of channel effectiveness
but not how that effectiveness varies with spending (saturation) or
persists over time (adstock).

\textbf{Meta Robyn} (\citeproc{ref-robyn2023}{Runge et al. 2023}) treats
calibration as a third objective in multi-objective optimization,
penalizing deviation from experimental lift estimates alongside fit and
budget allocation criteria. This provides a mechanism for steering model
selection toward experimentally-consistent regions of the parameter
space. However, without modeling the temporal structure of experimental
outcomes, the method cannot separately identify adstock and saturation
parameters---it constrains the aggregate effect without informing its
decomposition.

\textbf{pymc-marketing} (\citeproc{ref-orduz2024pymc}{Orduz 2024}) adds
adds lift test observations directly to the model likelihood, treating
experimental ROAS estimates as points on the saturation curve. This
principled Bayesian approach can inform saturation parameters when
experiments at different spending levels are available. However, because
it uses aggregate lift rather than time-series outcomes, it cannot
identify adstock dynamics---how long marketing effects persist after
spending stops.

Our structural estimation approach addresses these limitations directly.
By estimating the differencing equation on weekly time-series data, we
use: - The \emph{rate} of outcome gap emergence (when spending stops in
treatment) and closure (when spending resumes) to identify adstock decay
- The \emph{relationship} between spending differences and outcome
differences to identify saturation curvature\\
- The \emph{level} of outcome differences to identify effectiveness

All three parameters emerge from a single, theoretically-grounded
estimation rather than from separate heuristics or aggregate
constraints.

\subsection{Integration with Full MMM
Estimation}\label{integration-with-full-mmm-estimation}

This paper focuses on structural estimation as a standalone approach for
a single channel. Several strategies exist for integrating these
estimates into a full Marketing Mix Model covering multiple channels:

\textbf{Informative priors.} Use the posterior from structural
estimation as the prior for the corresponding channel in MMM estimation.
This requires specifying a joint prior that captures correlations
between parameters.

\textbf{Joint likelihood.} Incorporate the geo-test differencing
equation directly into the MMM likelihood, treating both observational
time-series and geo-test observations as arising from the same
underlying parameters. This approach mirrors how pymc-marketing
incorporates lift test observations into their Bayesian model---but
extended to use the full differencing equation rather than aggregate
ROAS estimates. The existing pymc-marketing codebase provides a natural
foundation for implementing this integration.

\textbf{Two-stage estimation.} Estimate channel parameters from
structural estimation, compute that channel's contribution to outcomes,
subtract from total sales, and estimate the remaining MMM on
residualized outcomes. This sequential approach is computationally
simpler but does not propagate uncertainty from the first stage.

\subsection{Limitations}\label{limitations}

Several limitations merit discussion, though most are shared with MMM
estimation more broadly rather than specific to our approach.

\textbf{Functional form assumptions.} Our approach assumes the MMM
specification---specifically, geometric adstock and logistic
saturation---correctly describes the true data-generating process. This
is not a limitation unique to our method; any MMM estimation requires
such assumptions. Indeed, our approach offers an advantage: because the
differencing equation contains only the parameters of interest after
differencing eliminates confounders, we can compare functional forms
(geometric vs.~delayed adstock, logistic vs.~Hill saturation) using
standard model selection criteria. This principled approach to
specification testing is not available in observational MMM estimation,
where confounding makes it difficult to distinguish model
misspecification from omitted variable bias.

\textbf{Geographic spillovers.} The method assumes no spillovers between
treatment and control areas. If advertising in control areas reaches
treatment consumers (e.g., as consumers travel between treatment and
control areas), the outcome difference will understate the true
treatment effect. This limitation is common to all geo-experiments, not
specific to our estimation approach.

\textbf{Parameter stability.} We assume parameters remain constant
across experimental periods. If marketing effectiveness varies
substantially over time---due to creative fatigue, competitive dynamics,
or changing consumer preferences---estimates reflect some form of
average effect across the experimental windows. This assumption is
standard in MMM more broadly; methods for accommodating time-varying
parameters (see Dew et al. (\citeproc{ref-dew2024mmm}{2024})) remain an
area for future extension of our approach.

\textbf{Experimental requirements.} The method requires geo-experiments
with sufficient temporal variation to identify dynamic parameters. Short
experiments (e.g., two-week tests) may not generate enough adstock
variation to precisely identify decay rates. Similarly, experiments must
span different spending levels to identify saturation curvature. These
requirements may exceed what standard geo-test designs provide, though
the informational gains justify enhanced experimental protocols.

Relatedly, the experiments identify the response curve only over the
range of spending they probe (fig.~\ref{fig:response}); recommendations
far outside that range---for example, a large budget expansion---lean on
the parametric form of the saturation function rather than on
experimental variation, and should be treated accordingly.

\section{Conclusion}\label{conclusion}

We have presented a structural estimation approach for recovering
Marketing Mix Model parameters from geo-experimental time-series. By
estimating the differencing equation directly---rather than reducing
experiments to point ROAS estimates---our method extracts substantially
more information from experimental investments than current practice
allows.

The approach contributes along several dimensions:

\begin{enumerate}
\def\labelenumi{\arabic{enumi}.}
\item
  \textbf{Causal identification of structural parameters.} While the
  observational MMM produces biased estimates due to endogenous
  budget-setting---even under oracle controls---our method recovers the
  true ROAS and response curve, together with credible estimates of
  adstock decay, saturation, and effectiveness, through experimental
  randomization.
\item
  \textbf{Full utilization of experimental data.} Existing calibration
  approaches reduce geo-experiments to aggregate statistics, discarding
  the temporal dynamics that identify how effects persist and diminish.
  Our method uses the full time-series, jointly identifying all three
  structural parameters from the same data.
\item
  \textbf{Efficient pooling across experiments.} Multiple
  geo-experiments for a single channel can be naturally combined by
  stacking observations of the same differencing equation, with
  precision improving as more experimental evidence accumulates.
\item
  \textbf{Principled functional form selection.} Because confounders are
  eliminated through differencing, alternative specifications (different
  adstock and saturation functions) can be compared using standard model
  selection criteria---an approach unavailable in observational
  estimation.
\end{enumerate}

For practitioners, this framework offers a method to extract greater
value from geo-experimental investments. Rather than obtaining only
point estimates of channel ROAS, the same experimental data can
calibrate the complete functional form needed for optimal budget
allocation across spending levels and time horizons.

The notebooks and other companion materials are available in the
\href{https://www.github.com/niklas-heusch/mmm-materials}{mmm-materials
GitHub repository}.

\section*{References}\label{references}
\addcontentsline{toc}{section}{References}

\protect\phantomsection\label{refs}

\clearpage
\appendices

\section{The MMM Identification Problem}\label{appendix-the-mmm-identification-problem}

This appendix provides a formal treatment of the identification problem
in Marketing Mix Models and how the differencing estimator resolves it.

\subsection{The Full MMM
Specification}\label{the-full-mmm-specification}

The standard Marketing Mix Model specifies outcomes as a function of
marketing spending across \(M\) channels plus control variables:

\[
y_t = \beta_0 + \sum_{m=1}^{M} \beta_m \cdot \text{sat}(x^*_{t,m}; \lambda_m) + \sum_{c=1}^{C} \gamma_c \, z_{t,c} + \sum_{u=1}^{U} \delta_u \, w_{t,u} + \varepsilon_t
\]

where:

\begin{itemize}
\item
  \(y_t\) is the outcome (e.g.~revenue) in period \(t\)
\item
  \(x^*_{t,m}\) is the adstock-transformed spend for channel \(m\), with
  decay parameter \(\alpha_m\)
\item
  \(\text{sat}(\cdot; \lambda_m)\) is the saturation function for
  channel \(m\)
\item
  \(z_{t,c}\) are observed control variables (seasonality indicators,
  pricing, economic conditions)
\item
  \(w_{t,u}\) are unobserved factors affecting demand (competitor
  actions, macroeconomic sentiment, etc.)
\item
  \(\varepsilon_t\) is idiosyncratic noise
\item
  Each channel has three parameters: \(\alpha_m\) (adstock decay),
  \(\lambda_m\) (saturation), \(\beta_m\) (effectiveness)
\end{itemize}

\subsection{The Identification
Problem}\label{the-identification-problem}

For the MMM to yield causal estimates of marketing effectiveness, the
following conditional independence assumption must hold:

\[
\varepsilon_t \perp x_{t,m} \mid z_t \quad \text{for all } m
\]

That is, the error term must be independent of marketing spending
conditional on the observed controls. This requires that \(z_t\)
captures \emph{all} factors that simultaneously affect both marketing
decisions and outcomes.

In practice, this assumption is virtually impossible to satisfy:

\textbf{Strategic timing.} Marketing teams plan campaigns around
anticipated demand. A retailer increases TV spending before Black Friday
because sales will be high---not because the ads will cause the sales.
Without controlling for ``anticipated holiday demand,'' the model
conflates seasonal effects with advertising effects.

\textbf{Algorithmic optimization.} Automated bidding systems (e.g.,
Google Smart Bidding) increase spending when conversion rates are high.
But high conversion rates may reflect strong underlying demand rather
than marketing effectiveness. The algorithm's response to demand signals
creates correlation between \(x_t\) and \(\varepsilon_t\).

\textbf{Budget responses.} When sales exceed expectations in one
quarter, marketing budgets often increase in the next. This creates
correlation between past outcomes (which include past \(\varepsilon\))
and current spending decisions.

\textbf{Unobservable factors.} Even with extensive controls, factors
like competitor actions, word-of-mouth momentum, or macroeconomic
sentiment affect both outcomes and marketing decisions in ways that
cannot be fully captured.

The consequence is bias: MMM estimates reflect some unknown mixture of
true causal effects and spurious correlation from endogenous
budget-setting.

\subsection{How Differencing Resolves
Identification}\label{how-differencing-resolves-identification}

The differencing estimator achieves identification through a
fundamentally different mechanism: experimental randomization rather
than conditional independence.

In a geo-experiment testing channel \(A\), we can write outcomes for
treatment and control groups as:

\[
y_t^{\text{control}} = \mu_t + \beta_A \cdot \text{sat}(x^{*,\text{control}}_{t,A}; \lambda_A) + \varepsilon_t^{\text{control}}
\]

\[
y_t^{\text{treatment}} = \mu_t + \beta_A \cdot \text{sat}(x^{*,\text{treatment}}_{t,A}; \lambda_A) + \varepsilon_t^{\text{treatment}}
\]

where
\(\mu_t = \beta_0 + \sum_{m \neq A} \beta_m \cdot \text{sat}(x^*_{t,m}; \lambda_m) + \sum_c \gamma_c z_{t,c} + \sum_u \delta_u w_{t,u}\)
contains all common factors.

The crucial observation is that \(\mu_t\) is \textbf{identical} across
treatment and control by construction: - Random assignment ensures the
groups are statistically equivalent at baseline - Non-tested channels
have identical spending in both groups - Observed controls (seasonality,
pricing) take the same values - Unobserved factors \(w_{t,u}\) are
balanced by randomization

Taking differences:

\[
\begin{aligned}
y_t^{\text{control}} - y_t^{\text{treatment}} &= \beta_A \bigl[ \text{sat}(x^{*,\text{control}}_{t,A}; \lambda_A)  \\
&\quad 
- \text{sat}(x^{*,\text{treatment}}_{t,A}; \lambda_A) \bigr] + \nu_t
\end{aligned}
\]

The term \(\mu_t\) vanishes. The remaining error
\(\nu_t = \varepsilon_t^{\text{control}} - \varepsilon_t^{\text{treatment}}\)
is the difference in idiosyncratic shocks, which is uncorrelated with
the spending difference by randomization.

\textbf{No conditional independence required.} We do not need to assume
that controls capture all confounders. We do not need to model how
budgets are set. Randomization handles all confounding---observed and
unobserved---automatically.

\textbf{Complete parameter identification.} The differencing equation
identifies all three structural parameters (\(\alpha_A\), \(\lambda_A\),
\(\beta_A\)) from observable quantities: outcome differences and
spending in both groups.

\subsection{Relationship to Standard Econometric
Approaches}\label{relationship-to-standard-econometric-approaches}

The differencing estimator can be viewed through several econometric
lenses:

\textbf{Difference-in-differences.} Like DiD, we compare treatment and
control over time. Unlike standard DiD, we estimate structural
parameters rather than average treatment effects, and we use the full
time-series rather than pre/post comparisons.

\textbf{Instrumental variables.} Randomization provides an instrument
(treatment assignment) for spending. However, our approach estimates the
structural form directly rather than using 2SLS.

\textbf{Potential outcomes.} The parallel universes framing connects to
the Rubin causal model. Treatment and control represent draws from the
same potential outcome distribution, differing only in the realized
treatment.

\section{Implementation Details}\label{appendix-implementation-details}

This appendix summarizes the probabilistic model and key implementation
details. Complete Python code implementing all methods is available in
the companion Jupyter notebook.

\subsection{Probabilistic Model in Plate
Notation}\label{probabilistic-model-in-plate-notation}

We estimate the differencing equation using Bayesian inference in PyMC.
The model structure is shown in plate notation below, generated directly
from PyMC's \texttt{model\_to\_graphviz} function:

\begin{figure}
\centering
\pandocbounded{\includegraphics[keepaspectratio]{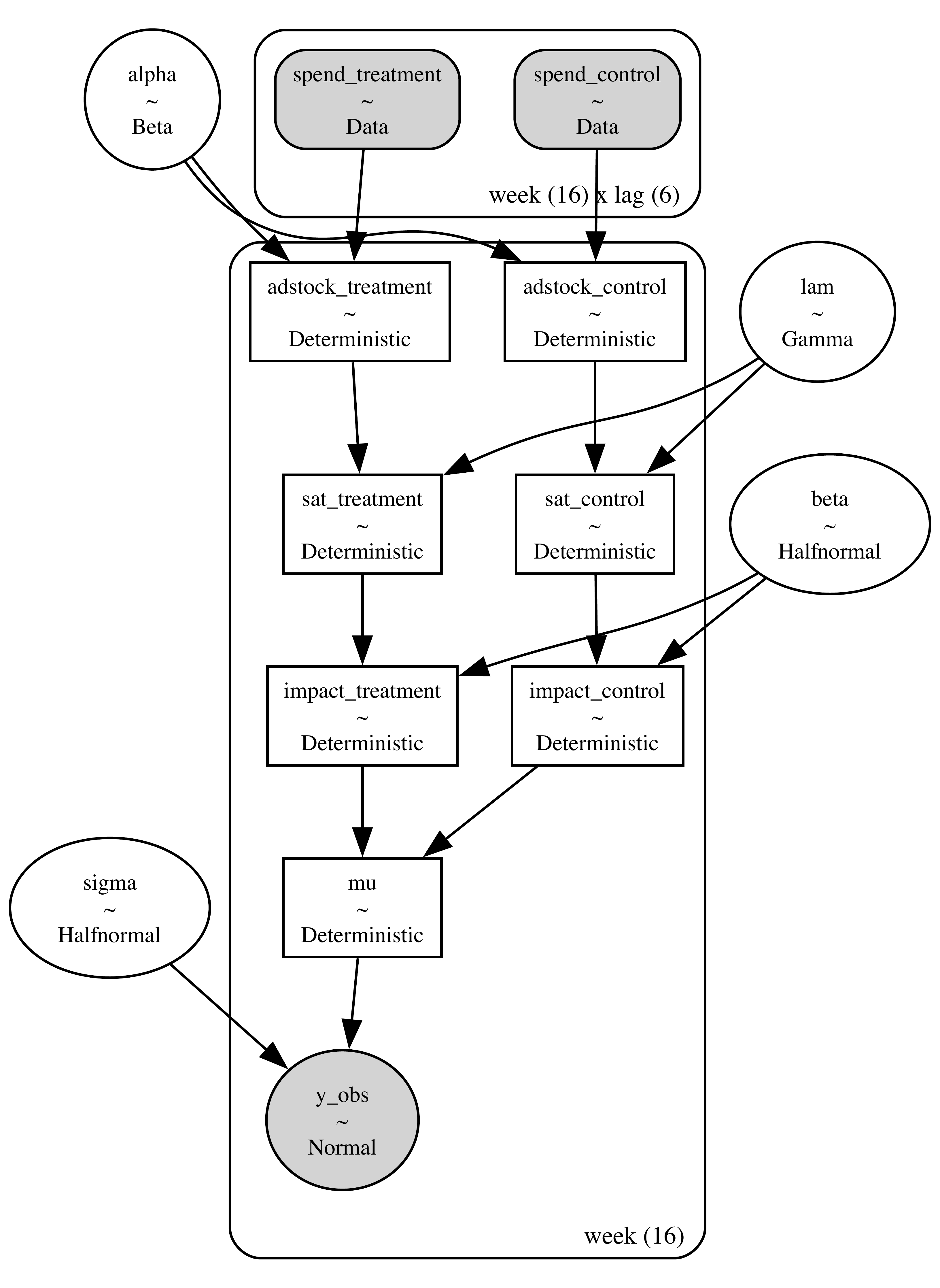}}
\caption{Structural Estimation Model: Plate
Notation}\label{fig:platenotation}
\end{figure}

\textbf{Reading the plate notation:}

The diagram shows the complete data flow for both treatment and control
groups:

\begin{enumerate}
\def\labelenumi{\arabic{enumi}.}
\item
  \textbf{Observed inputs} (rectangular nodes): Marketing spends for
  control (\(x^{\text{ctrl}}\)) and treatment (\(x^{\text{treat}}\))
  groups, each containing \(L\) weeks of data (current + lagged spends).
\item
  \textbf{Shared parameters} (elliptical nodes at top): The key
  structural parameters---\(\alpha\) (adstock decay), \(\lambda\)
  (saturation), and \(\beta\) (effectiveness)---are \emph{shared}
  between both groups. This encodes the assumption that the marketing
  channel works the same way regardless of experimental group
  assignment.
\item
  \textbf{Parallel transformations} (deterministic nodes): For
  \emph{each} group:

  \begin{itemize}
  \tightlist
  \item
    \textbf{Adstock}:
    \(x^*_t = \sum_{\ell=0}^{L-1} \alpha^\ell x_{t-\ell} / \sum_{\ell=0}^{L-1} \alpha^\ell\)
  \item
    \textbf{Saturation}: \(\text{sat}(x^*_t; \lambda)\)
  \item
    \textbf{Impact}: \(\beta \cdot \text{sat}(x^*_t; \lambda)\)
  \end{itemize}
\item
  \textbf{Differencing} (\(\mu\)): The predicted outcome difference
  equals \(\text{impact}_{\text{ctrl}} - \text{impact}_{\text{treat}}\).
\item
  \textbf{Likelihood} (\(\Delta y\), shaded): Observed outcome
  differences follow \(\mathcal{N}(\mu, \sigma)\).
\end{enumerate}

The plate (rectangle around observation-level nodes) indicates that
transformations are computed for each of the \(T\) time periods.
Crucially, the parameters sit \emph{outside} the plate---they are
estimated once and applied identically to both groups across all
observations.

\subsection{Computational Notes}\label{computational-notes}

We use PyMC's No-U-Turn Sampler (NUTS) with 4 chains, 1000 tuning and
2000 sampling iterations each, and target acceptance rate 0.95 (1000
sampling iterations for the observational MMM benchmarks). Convergence
is assessed via R-hat \textless{} 1.01 and ESS \textgreater{} 400; all
runs pass. Estimation completes in under 30 seconds on a modern laptop.
The observational MMM uses the same media-parameter priors as the
structural model (the priors table of Section 4); its remaining priors
are Normal(0, 0.3) on control coefficients, Normal(0, 0.2) on Fourier
coefficients, Normal(0.5, 0.2) on the intercept, and HalfNormal(0.1) on
the noise scale, with Normal(0, 2) on the oracle specification's control
coefficients so that priors cannot be accused of holding perfect
covariates back.

\section{Additional Results}\label{appendix-additional-results}

\subsection{Sensitivity to Prior
Specification}\label{sensitivity-to-prior-specification}

We examined sensitivity to prior choices by re-running the structural
estimation with alternative priors:

\begin{table}[!htbp]
\centering
\footnotesize
\caption{Sensitivity to prior specification}\label{tbl-prior-sensitivity}
\begin{tabular}{@{}lccc@{}}
\toprule
Prior Set & $\alpha$ & $\lambda$ & ROAS (4 tests) \\
\midrule
Baseline & Beta(1,3) & Gamma(3,1) & 4.14 [3.77, 4.51] \\
Diffuse & Beta(1,1) & Gamma(1,0.5) & 4.07 [3.69, 4.44] \\
Informative & Beta(2,8) & Gamma(5,2) & 4.14 [3.82, 4.45] \\
\bottomrule
\end{tabular}
\par\vspace{0.8em}
\begin{minipage}{\columnwidth}
\footnotesize
\textit{Note:} Prior distributions for $\alpha$ (adstock) and $\lambda$ (saturation). ROAS estimates with 90\% credible intervals.
\end{minipage}
\end{table}

Results are robust to reasonable prior variations: point estimates shift
by less than 2\%, and all credible intervals contain the true ROAS. The
same holds on the other side of the comparison: re-estimating the
realistic MMM under the alternative prior set used in the companion blog
series (Beta(2, 1) on carryover, Gamma(2, 1) on saturation, Normal(0,
0.5) on the media coefficients) leaves its bias intact (11.67×
vs.~10.66× under the baseline priors)---the endogeneity bias is
likelihood-driven, not prior-driven.

\end{document}